\documentclass[preprint2]{aastex}
\usepackage{aastexug}

\shorttitle{Li Abundances in bulgelike Stars}
\shortauthors{Pompeia et al.}

\begin{document}
\input{epsf}

\title{Detailed Analysis of Nearby Bulgelike Dwarf Stars II. Lithium Abundances}

\author{Luciana Pomp\'eia and Beatriz Barbuy}
\affil{Instituto Astron\^omico e Geof{\'i}sico, USP, 01060-970 S\~ao Paulo, Brazil }
\email{pompeia@iagusp.usp.br, barbuy@iagusp.usp.br}

\and
\author{Michel Grenon}
\affil{Observatoire de Gen\`eve, Chemin des Maillettes 51, CH-1290 Sauverny, Switzerland}
\email{Michel.Grenon@obs.unige.ch}

\and

\author{Bruno Vaz Castilho}
\affil{Laborat\'orio Nacional de Astrof\'{\i}sica, R. Estados Unidos, 154, 37504-364,
Itajub\'a - MG, Brazil}
\email{bruno@lna.br}


\begin{abstract}

Li abundances are derived for a sample of bulgelike stars with isochronal ages of 
10-11 Gyr. These stars have orbits with pericentric distances, R$_{p}$, as small as 
2-3 kpc and Z$_{max}$ $<$ 1 kpc. The sample comprises G and K dwarf stars in the 
metallicity range -0.80$\leq$[Fe/H]$\leq$ +0.40. Few data of Li abundances in old turn-off 
stars ($\geq$ 4.5 Gyr) within the present metallicity range are available. M67 (4.7 Gyr) and 
NGC 188 (6 Gyr) are the oldest studied metal-rich open clusters with late-type stars. Li 
abundances have also been studied for few samples of old metal-rich field stars. In the 
present work a high dispersion in Li abundances is found for bulgelike stars for 
all the metallicity range, comparable with values in M67. The role of metallicity and age 
on a Li depletion pattern is discussed. The possible connection between Li depletion and oxygen 
abundance due to atmospheric opacity effects is investigated.
 
\end{abstract}
\keywords{stars: abundances - stars: chemical evolution - stars: late-type - 
elments: lithium}


\section{Introduction}

Lithium is a key element in astrophysics. The inference of the original lithium abundance in 
the universe should provide the cosmic ratio of baryons to photons at primordial times, 
and constrain the standard model of Big Bang nucleosynthesis. 
Li is destroyed in stars by $^{7}$Li(p,$\alpha$)$^{4}$He reactions at T $\geq$ 2.5 
$\times$ 10$^{6}$ K. On the other hand, different sources of Li, besides the primordial 
nucleosynthesis, have been proposed: novae, asymptotic and red giant branch stars, 
C-stars and Type II SNe, and spallation of C, N, O elements by galactic cosmic rays in the 
interestellar medium. To infer the primordial Li abundance an overall understanding of the 
different destruction/production mechanisms and their rates is required.   

In a classical work, \citet{spi82} detected an almost uniform Li abundance in halo 
stars. This "Li plateau", with A(Li)\footnote{A(Li) = log $\varepsilon$(Li)= log 
[N(Li)/N(H)] + 12} $\sim$ 2.1, is constant  for all metal-poor 
stars within the temperature range 5700 K $\leq$ T$_{\rm eff}$ $\leq$ 6300 K. 
Many authors consider this value as the primordial Li abundance of the protogalactic cloud. 
Others claim that it is depleted by about 0.2-0.3 dex from the primordial value (for recent 
reviews see Cayrel 1998; Spite et al. 1998; Pinsonneault et al. 2000).

Solar-system observations led to meteoritic Li abundances of A(Li) $\sim$ 3.3 
(Grevesse et al. 1996). T Tauri stars provided a similar value, A(Li) $\sim$ 3.2 {\citep{mag92}
and pre-MS stars in the Orion Nebula provided values of A(Li) $\sim$ 3.6 
{\citep{kin93}} and A(Li) $\sim$ 3.2 {\citep{cun95}}.} 
These observations show that there is a roughly constant galactic interestellar 
medium (ISM) abundance. If the Li plateau value is the actual primordial value, 
the ISM abundance is enhanced by a factor of ten. Therefore the understanding of 
the galactic Li history demands a complete theory of stellar evolution and their 
contribution to the ISM enrichment. 

Standard models, which have successfully explained stellar evolution and HR diagrams, 
predict that convection is the only mechanism that rules Li depletion in low-mass 
stars. In these models Li depletion is a function of stellar mass, metallicity and 
age (e.g. D'Antona \& Mazzitelli 1984; Proffitt \& Michaud 1989). 

Open clusters are natural targets to probe these models because they 
have stellar contents of high metallicity sampling an evolutionary sequence. However, 
their study revealed a more complex picture than that outlined by standard models. 
MS depletion is expected to occur only in the lower mass (M$<$0.9 M$_{\odot})$ stars. 
But observations indicate that a depletion mechanism acts during the entire MS 
lifetime even in stars where the temperature at the bottom of the convective zone (CZ) 
is not high enough to burn Li. In addition, stars with the same age, composition and mass 
show dispersions in A(Li) as high as 1.5 dex (e.g. Soderblom et al. 1993a for Pleiades 
late-G and early-K stars, and Pasquini et al. 1997 for M67).

Non-standard models with different depletion mechanisms have been suggested to account 
for the observations: mixing driven by angular momentum loss, microscopic 
diffusion, internal wave diffusion, differential rotation and depletion by MS mass 
loss (see Pinsonneault 1997, Deliyannis et al. 2000 and Pinsonneault et al. 2000). 
None of these models provided a satisfactory fit to observations alone, 
indicating that two or more mechanisms are acting together.

A comprehensive literature is available for lithium in open clusters. 
In the last two decades, several groups observed and derived Li abundances for several
open clusters solar-type stars. Some examples of studied young clusters are: Blanco 1 
{\citep{jef97}}, $\alpha$ Per {\citep{boe88,lop94,bal96,ran97}}, Ursa Major Group {\citep{so93b}} 
and Pleiades {\citep{boe88,lop94,so93a,jon96}}. With ages in the range 30-100 Myr, they have 
stars which have just arrived to the ZAMS. Therefore their abundance patterns are records of PMS 
depletion activity. Large spreads in Li abundance are observed in these clusters in 
the temperature range 5500 K $\leq$ T$_{\rm eff}$ $\leq$ 4500 K.

M34 (Jones et al. 1997), NGC 6475 (James et al. 1997), with 200-300 Myr and NGC 6633 
\citep{jef97}, Hyades (Thornburn, 1993, Soderblom et al. 1995) and Praesepe (Soderblom 
et al. 1993c) with 600-800 Myr are young clusters with stars of relatively short 
evolution time spent in MS. NGC 754 \citep{hob86} and IC 4651 and NGC 3688 \citep{ran00} with $\sim$ 
2 Gyr are intermediate age clusters. These groups show decreasing Li abundances with age.

Few data are available for old stars in open clusters or in the field for the present metallicity
(-0.80$\leq$[Fe/H]$\leq$ +0.40) and temperature (4700 K $\leq {\rm T_{eff}} \leq$ 5900 K) ranges. M67 
(Pasquini et al. 1997; Jones et al. 1999) and NGC 188 {\citep{hob88}} with ages 4.7 and 6 Gyr 
respectively, are the oldest open clusters with derived Li abundances. Field samples with inhomogeneous 
ages and kinematics were also studied \citep{pas94,fav96,che01}. For hotter stars the behavior of 
A(Li) vs. temperature was studied by e.g. \citet{lam91} and \citet{boe01}.

In present work we derive Li abundances for a kinematically selected sample. The 
stars from this sample have highly eccentric orbits indicating an inner disk or 
bulge origin. The sample selection and kinematical properties are described in 
Grenon (1999, 2000). A detailed analysis of these stars was presented in Pomp\'eia, 
Barbuy \& Grenon (2001, hereafter Paper I). Binaries are rejected using Hipparcos 
photometry and radial velocity measurements. Isochronal ages are 10-11 Gyr, 
making this one of the oldest samples with derived Li abundances in the studied metallicity 
range.


\section{Observations and Analysis}

The spectra were obtained in September 1999, at the 1.52m telescope of ESO, 
La Silla, with the FEROS spectrograph. The standard star+sky configuration was used. 
The spectral coverage is from 356 to 920 nm, with a R=48,000 resolution. 
Data reduction was performed using the ESO pipeline package for reductions of FEROS 
data (DRS), in MIDAS environment.

Stellar parameters are those derived in Paper I, where effective temperatures were 
derived using H$\alpha$ profiles and surface gravities were inferred by requiring 
ionization equilibrium of Fe I and Fe II lines. MARCS model atmospheres (Gustafsson 
et al. 1975) were employed. Metallicities and microturbulent velocities were determined 
by using curves of growth for Fe I and Fe II. Stellar masses were derived from isochrones 
of Vandenberg (1985) and  VandenBerg \& Laskarides (1987). A detailed description of the 
determination of stellar parameters is presented in Paper I. In Table 1 effective 
temperatures, gravities, microturbulence velocities, metallicities   
and masses are reported.    


\subsection{Li Abundances}

Li abundances were determined by using synthetic spectra in the region of the
Li doublet at $\lambda$ 6707.76 ${\rm \AA}$. The spectrum synthesis code is described 
in Cayrel et al. (1991). The atomic line list in the region is reproduced in
Table 2 and molecular lines of TiO, C$_{2}$ and CN are included. 

\placetable{tb2}

The $\lambda$ 6103.4 ${\rm \AA}$ Li line is known to be less perturbed by 
NLTE effects than the resonance one at $\lambda$ 6707.76 ${\rm \AA}$. However due 
to the metallicity and atmospheric parameters range of our sample, it is heavily 
blended and undetectable for most of the stars. The NLTE effects for the $\lambda$ 6707.76 
${\rm \AA}$ doublet of our sample stars are of the order of [Li/H] $\sim$ 0.012 
\citep{car94} and do not affect the results.

Errors in Li abundances are dominated by temperature uncertainties. The estimated 
error in A(Li) is 0.07 dex for a T$_{\rm eff}$ change of 100 K. The calculated Li 
abundances are also reported in Table 1. In Fig. 1 the Li line syntheses for HD 211706 
and HD 10576 are shown. 

\begin{figure}
\label{fg1}
\plotone{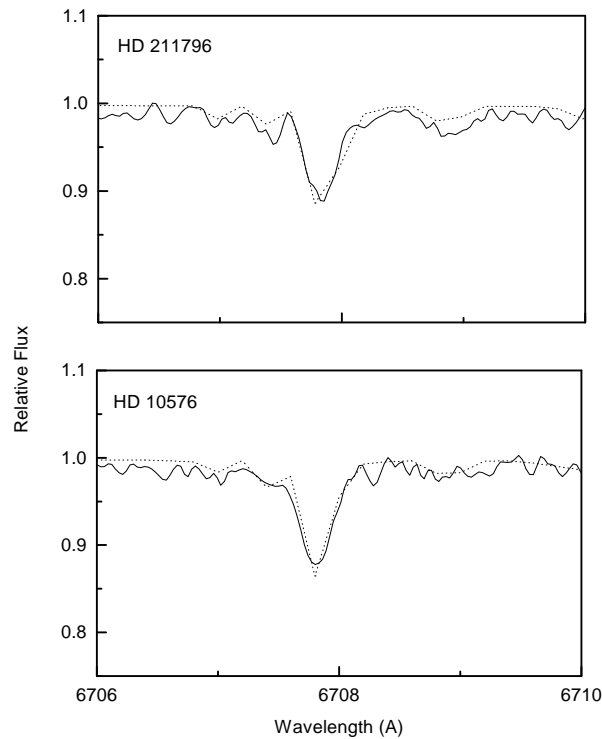}
\caption{Synthetic (dashed line) and observed (solid line) spectra of HD 211706 and HD 10576.}
\end{figure}

\placefigure{fg1}


\section{Discussion}

\subsection{Li vs. T$_{\rm eff}$}

In Fig. 2 Li abundances vs. T$_{\rm eff}$ are plotted. Most of the determined abundances
for the sample stars represent upper limits indicating that the line depth is below the 
limit of 2$\sigma$ of the noise. Nevertheless, an upper envelope of high-Li stars is also observed. 

A large spread in Li abundances is present for stars with the same temperature. Large spreads in 
Li abundances were also inferred for other samples of turn-off field stars with different 
compositions and ages \citep{lam91,pas94,fav96,che01}, and even for very homogeneous 
samples as M67 G-type stars \citep{pas97,jon99}.

\begin{figure}
\label{fg2}
\plotone{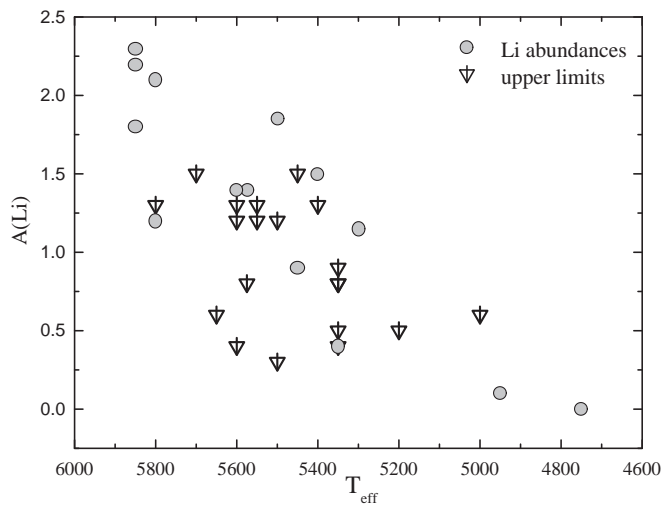}
\caption{Li abundances vs. effective temperatures for the sample stars. Down triangles
represent upper limits.}
\end{figure}

\placefigure{fg2}
 
\subsection{Li vs. age}

The correlation between Li abundances and age is examined in Fig. 3 where the 
A(Li) vs. T$_{\rm eff}$ is plotted compared to that for M67. As shown in this 
figure, the two samples overlap. The lack of a depleted pattern of 
bulgelike relative to M67 stars suggest that Li depletion mechanisms become inefficient 
with age. Open clusters observations are in agreement to this suggestion. The depletion rate 
apparently decreases with time given that higher depletion is observed among young clusters with different
ages (50 to 600 Myr) than among old clusters (1 to 4 Gyr) \citep{jef00}. \citet{pas00} suggested 
that no depletion mechanism acts for ages older than $\sim$ 1.6 Gyr. Based on a sample of field stars, 
\citet{che01} have also claimed that Li depletion occurs early in life, at ages $\leq$ 1.5 Gyr. Our sample, 
with much older stars and with some high-Li, supports this suggestion.

\begin{figure}
\label{fg3}
\plotone{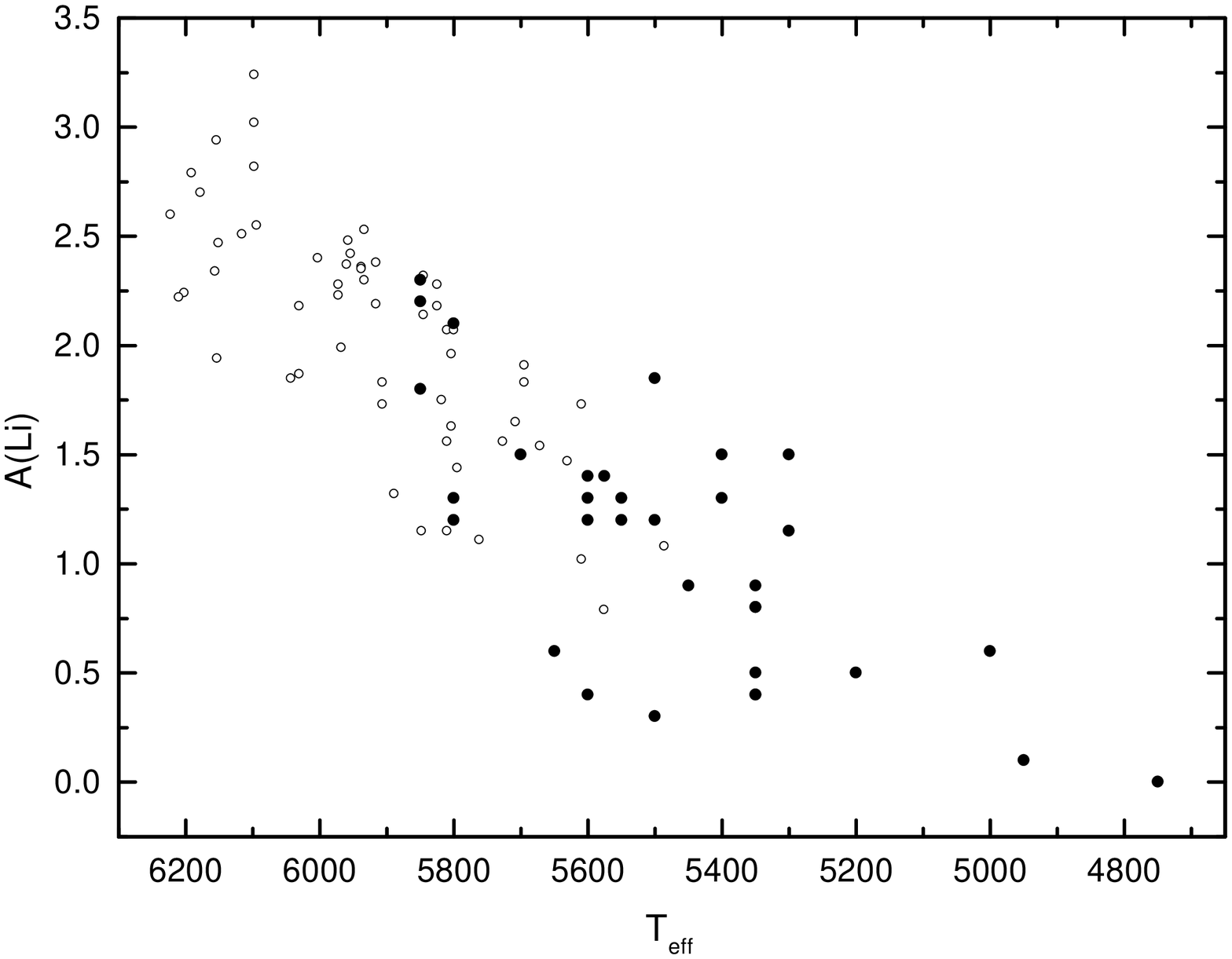}
\caption{Li abundances vs. effective temperatures for our sample stars (solid circles) 
compared to M67 cluster stars (open circles) \citep{pas97}.}
\end{figure}

\placefigure{fg3}      

\subsection{Li vs. Mass}     
           
Standard models predict that high-mass stars preserve more of their Li than lower mass
stars. In Fig. 4 we plotted the A(Li) vs. Mass (M/M$_{\odot}$). A slight trend 
of Li abundance with mass is observed although with some scatter, probably due to
the action of other depletion mechanisms, such as rotation-driven mechanisms which 
depend on the rotation history of the star.

\begin{figure}
\label{fg4}
\plotone{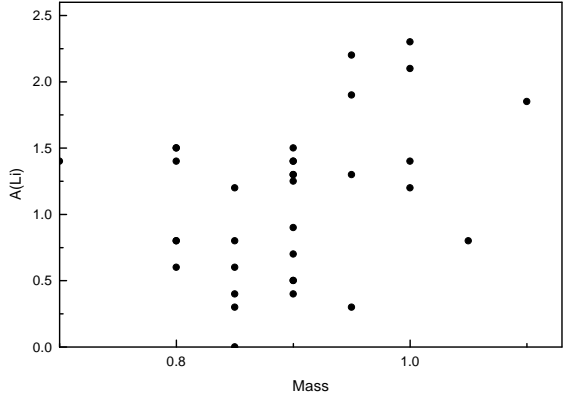}
\caption{Li abundances vs. Mass (M/M$_{\odot}$).}
\end{figure}

\placefigure{fg4}
           
\subsection{Li vs. [Fe/H]}

The depth of the convection zone is larger for higher metallicity stars, therefore these
stars are predicted to burn more their Li than lower metallicity stars with the same 
temperature. In order to check the role of metallicities in Li depletion we compare in Fig. 5 
our Li data to that for NGC 6397 (Castilho et al. 2000), a 11.5 Gyr cluster (Chaboyer 1998) 
with metallicity [Fe/H] = -2.0. This figure shows that, although older, NGC 6397 stars have 
preserved more of their Li content than most of the bulgelike stars with the same temperature. 
On the other hand, some high-Li bulgelike stars are also observed.

In order to test the correlation between metallicity and the Li content we plotted A(Li) vs. [Fe/H]
in Fig. 6. Different symbols represent different ranges of effective temperature.   
No apparent trend between Li abundance and metallicity is found for any of the temperature 
ranges. Nevertheless, a weak correlation with metallicity may be present contributing to the
dispersion in Li abundances.

\begin{figure}
\label{fg5}
\plotone{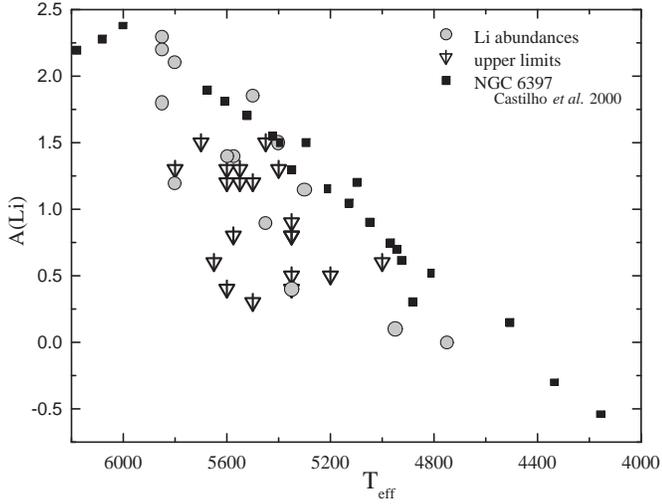}
\caption{Li abundances vs. effective temperatures for our sample stars and for 
NGC 6397 \citep{cas00}.}
\end{figure}

\placefigure{fg5}

\subsection{Li vs. [O/H]}

Swenson et al. (1994) developed stellar models taking into account improved interior and surface
opacities. They argue that elements such as Si, O, Ne and Mg may induce opacity changes, resulting 
in different Li burning rates. Their results pointed out that oxygen, together with iron, is the main 
contributor to the opacity. They estimated that an oxygen enhancement of [O/Fe] $\sim$ +0.20 could 
increase significantly the Li depletion rate; using new opacity tables with enhanced oxygen-to-iron 
ratio they reproduced the observed pattern of the Hyades stars. 

In order to analyze if oxygen abundances contribute to Li depletion we compared Li 
abundances with the oxygen abundances derived in Paper I. In Fig. 7 A(Li) vs. [O/H] 
are plotted. A statistical analysis is performed and a correlation coefficient of 
$r$ = - 0.23 is inferred. This coefficient indicates that essentially no dependence  
exists between Li and oxygen abundances.  
 
\begin{figure}
\label{fg6}
\plotone{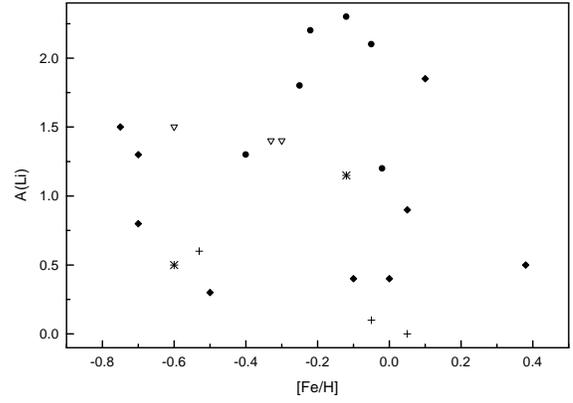}
\caption{Li abundances vs. metallicity. The symbols represent different 
temperature ranges: 4850-5000 K (stars), 5050-5300 K (crosses), 5350-5500 K (down triangles),
 5550-5700 K (diamonds), 5750-5900 K (circles)}.
\end{figure}

\placefigure{fg6}

\section{Summary}

In the present work lithium abundances for a sample of bulgelike stars with isochronal 
ages of 10-11 Gyr are reported. High-Li and low-Li stars are observed. The overlap between 
lithium curves for M67, a 4 Gyr open cluster, and our bulgelike sample, $\sim$ 5 Gyr older,  
seems to indicate that the depletion mechanisms become inefficient with age and no depletion
occurs during the latter stages (ages older than $\sim$ 1.5 Gyr) of the MS.

The derived Li abundances show that a high dispersion is present for bulgelike stars with 
the same temperature comparable to that observed in field samples and in M67 stars. 
No apparent correlation between [Fe/H] and A(Li) is found, although differences in 
metallicities may account for part of the observed Li dispersion. A spread in the Li 
abundance vs. stellar mass with a possible weak trend is obtained.

\bigskip

$Acknowledgments$  L. P. acknowledges the FAPESP PhD fellowship n$^{\rm o}$ 
98/00014-0. We acknowledge FAPESP project n$^{\rm o}$ 1998/10138-8.



\begin{deluxetable}{crrrrrr}
\label{tb1}
\tablecolumns{7}
\tablewidth{0pc}
\tablecaption{Atmospheric Parameters and Lithium abundances for the sample stars}
\tablewidth{0pt}
\tablehead{
\colhead{Name} & \colhead{T$_{\rm eff}$} & \colhead{M/M$_{\odot}$} & \colhead{log g} & 
\colhead{[Fe/H]} & \colhead{$\xi(\rm kms^{-1})$} & \colhead{A(Li)} 
}
\startdata
HD 143016  &  5575  &  0.85 & 3.8  &  -0.50  &  1.0  &  $<$ 0.8 \\
HD 143102  &  5500  &  1.10 & 3.7  &  0.10  &  0.9  &  1.85 \\
HD 148530  &  5350  &  0.90 & 4.3  &  0.00  &  0.5  &  $<$ 0.4 \\
HD 149256  &  5350  &  0.80 & 3.6  &  0.26  &  1.1  &  $<$ 0.8 \\
HD 152391  &  5300  &  0.90 & 3.9  &  -0.12  &  0.9  &  1.25 \\
HDE326583  &  5600  &  1.00 & 3.7  &  -0.50  &  0.6  &  $<$ 1.2 \\
HD 175617  &  5550  &  0.80 & 4.7  &  -0.48  &  0.5  &  $<$ 1.4 \\
HD 178737  &  5575  &  0.90 & 4.0  &  -0.33  &  0.6  &  1.4 \\ 
HD 179764  &  5450  &  0.90 & 4.2  &  0.05  &  0.5  &  0.7 \\
HD 181234  &  5350  &  0.90 & 4.1  &  0.38  &  0.8  &  $<$ 0.5 \\
HD 184846  &  5600  &  0.85 & 4.0  &  -0.25  &  0.8  &  $<$ 0.4 \\
BD-176035  &  4750  &  0.85 & 3.8  &  0.05  &  1.0  &  0.0 \\
HD 198245  &  5650  &  0.80 & 4.3  &  -0.65  &  0.5  &  $<$ 0.6 \\
HD 201237  &  4950  &  0.95 & 4.3  &  -0.05  &  0.5  &  0.3 \\
HD 211276  &  5500  &  0.85 & 4.0  &  -0.55  &  0.5   &  $<$ 1.2 \\
HD 211532  &  5350  &  0.80 & 4.7  &  -0.70  &  0.5  &  $<$ 0.8 \\ 
HD 211706  &  5800  &  1.00 & 3.7  &  -0.05  &  1.0  &  2.1 \\
HD 214059  &  5550  &  0.90 & 3.8  &  -0.33  &  0.65  &  $<$ 1.3 \\
CD-4015036 &  5350  &  0.90 & 4.1 &  -0.10  &  0.5  &  0.5 \\
HD 219180  &  5400  &  0.80 & 4.4  &  -0.70  &  0.5  &  1.5 \\
HD 220536  &  5850  &  0.95 & 3.9  &  -0.22  &  1.0  &  2.2  \\
HD 220993  &  5600  &  0.90 & 4.0  &  -0.30  &  0.7  &  $<$ 1.4 \\
HD 224383  &  5800  &  1.00 & 4.1  &  -0.02  &  1.0  &  1.4 \\
HD   4308  &  5600  &  0.90 & 4.0  &  -0.40  &  0.7  &  $<$ 1.3 \\
HD   6734  &  5000  &  1.05 & 3.1  &  -0.53  &  0.8  &  $<$ 0.8 \\
HD   8638  &  5500  &  0.85 & 4.1  &  -0.50  &  0.9  &  $<$ 0.3 \\
HD   9424  &  5350  &  0.90 & 4.0  &  0.00   &  0.8  &  $<$ 0.9 \\
HD  10576  &  5850  &  1.00 & 3.6  &  -0.12  &  1.25  &  2.3 \\
HD  10785  &  5850  &  0.95 & 4.2  &  -0.25  &  1.0  &  1.9 \\ 
HD  11306  &  5200  &  0.85 & 4.3  &  -0.60  &  0.6  &  $<$ 0.6  \\
HD  11397  &  5400  &  0.80 & 4.0  &  -0.70  &  0.6  &  $<$ 1.5 \\
HD  14282  &  5800  &  0.95 & 3.7  &  -0.40  &  1.0  &  $<$ 1.3 \\
HD  16623  &  5700  &  0.90 & 4.0  &  -0.60  &  1.0  &  $<$ 1.5 \\
BD-02 603  &  5300  &  0.90  & 3.9  &  -0.75  &  0.8  &  $<$ 1.5 \\
HD  21543  &  5650  &  0.70 &  4.1  &  -0.55  &  0.5  &  $<$ 1.4 \\
\enddata
\end{deluxetable}

\begin{deluxetable}{crrrr}
\label{tb2}
\tabletypesize{\scriptsize}
\tablecolumns{5}
\tablewidth{0pc}
\tablecaption{The Line List and gf-values Around the Li Line}
\tablehead{
\colhead{Species} & \colhead{$\lambda$ (\rm \AA)} & \colhead{$\chi$(\rm eV)} & \colhead{log gf}  
& \colhead{References}
}
\startdata
Si1  & 6707.050 & 5.95  & -5.00 & B  \\ 
Fe1  & 6707.441 & 4.68  & -2.40 & B  \\
Sm2  & 6707.450 & 0.93  & -1.04 & L  \\   
V1   & 6707.563 & 2.74  & -1.53 & B  \\
Cr1  & 6707.644 & 4.21  & -2.14 & B  \\
Ce2  & 6707.740 & 0.50  & -3.02 & L  \\
Li7  & 6707.776 & 0.00  & 0.00  & S  \\
Li7  & 6707.927 & 0.00  & -0.30 & S  \\              
V1   & 6708.100 & 1.22  & -2.99 & B  \\
Fe1  & 6708.320 & 3.00  & -4.70 & B  \\
Ti1  & 6708.755 & 3.92  & -0.09 & B  \\ 
Fe1  & 6708.780 & 3.00  & -4.39 & B  \\
Fe1  & 6708.955 & 3.00  & -4.48 & B  \\
\enddata
\tablerefs{(B) \citet{bar99}, (L) \citet{lam93}, (S) \citet{spi82}}
\end{deluxetable}


\clearpage

\clearpage

\clearpage

\begin{figure}
\label{fg7}
\plotone{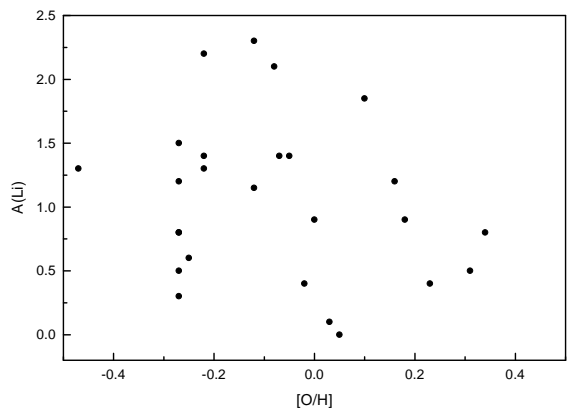}
\caption{Li abundances vs. oxygen abundances.}
\end{figure}

\end{document}